\documentclass[aps,prl,notitlepage]{revtex4-1}
\usepackage{graphics,graphicx,amssymb,subfig,array,txfonts,multirow,caption,verbatim,bm,bigstrut,color,psfrag,textcomp}
\begin{document}
\addtolength{\oddsidemargin}{-.13in}
\title{Design of a technique to measure the density of ultracold atoms in a short-period optical lattice in three dimensions with single atom sensitivity}
\author{M. D. Shotter}
\email{Email: mshotter@gmail.com}
\affiliation{National Institute of Standards and Technology,
100 Bureau Drive, Stop 8423,
Gaithersburg, MD 20899-8423, USA}
\altaffiliation{Previously: Department of Physics, University of Oxford, Oxford, UK OX1 3PU}
\date{24 November 2010}
\begin{abstract}
A measurement technique is described which has the potential to map the atomic site occupancies of ultracold atoms in a short-period three-dimensional optical lattice. The method uses accordion and pinning lattices, together with polarization gradient cooling and fluorescence detection, to measure the positions of individual atoms within the sample in three dimensions at a resolution of around half the atomic resonant wavelength. The method measures the site occupancy, rather than the parity of the site occupancy, of atoms in the lattice. It is expected that such measurements hold significant potential for the study of ultracold quantum dynamics.
\end{abstract}
\maketitle

In recent years there has been a surge of interest in the exploration of strongly interacting quantum systems of ultracold atoms in optical lattices. In addition to long coherence times, a high degree of control over the quantum Hamiltonian is possible, making such systems ideal for studies of many-body entanglement \cite{jaksch_cold_2005}. The direct measurement of the number of atoms at each site of the optical lattice provides a great deal of information about the system, particularly in terms of measurements of local correlation and excitation properties. The measurement of the positions of individual atoms within a strongly interacting quantum system in an optical lattice has been demonstrated very recently in two experiments \cite{bakr_probing_2010,sherson_single_2010}; it may be anticipated that the excellent results from these experiments will stimulate further research in this direction.

The present article outlines and models a measurement technique which potentially offers several advantages over the measurement techniques employed in the recent experiments \cite{bakr_probing_2010,sherson_single_2010}. The aim of the technique is to measure the number of atoms at each lattice site, for a dense three-dimensional sample of ultracold atoms held in a three-dimensional optical lattice formed by counter-propagating laser beams, for all lattice sites within a certain small volume (of order 10$\,\times\,$10$\,\times\,$10 lattice sites to a side).

The technique which will be outlined in this article, in common with the recent experiments \cite{bakr_probing_2010,sherson_single_2010}, uses fluorescence detection to measure the atomic density distribution. There are four major issues which must be overcome when using fluorescence detection to probe dense neutral atomic samples in an optical lattice on an atom-by-atom level. Firstly, each atom must be kept (to a high probability) within a small volume during the photon collection time in order to extract a strong and consistent spatially-localized signal. A deep optical lattice can be used to confine the atoms during the measurement, with the strong photon recoil heating compensated with polarization gradient cooling. This technique has been used in the recent experiments \cite{bakr_probing_2010,sherson_single_2010}, and is explored in detail in a separate article by the author \cite{shotter_large_2010}.

Secondly, the spatial resolution of the image should be high enough to resolve individual lattice sites. Strongly interacting dynamics occur preferentially in small-period lattices and at high atomic density, leading to substantial difficulties in resolving the sites; nevertheless this capability has been successfully demonstrated in the two recent experiments \cite{bakr_probing_2010,sherson_single_2010} using very high numerical aperture lens systems. 

Thirdly, exposure to intense resonant light during the measurement causes strong light-assisted collisional losses. The energy released in such collisions is such that the atoms are immediately ejected from the confining potential. These losses occur pairwise; isolated atoms are immune. The atoms are lost in a very small fraction of the measurement time, leaving no identifiable fluorescence signal. The current generation of resolved-atom experiments \cite{nelson_imaging_2007,bakr_quantum_2009,bakr_probing_2010,sherson_single_2010} measure the parity of the lattice site occupation numbers rather than the true number distribution. 

A fourth issue occurs when making measurements of a three-dimensional sample; such a measurement typically involves the tomographic reconstruction of the spatial distribution from a sequence of images at different depths. The measurement technique should be capable of distinguishing the signal from atoms at different depths within the sample; scattering from out-of-focus atoms should not overwhelm the signal from in-focus atoms.

This article outlines a technique which has the potential to resolve the latter three issues.

\begin{figure}
\subfloat[]{\includegraphics[scale=0.3]{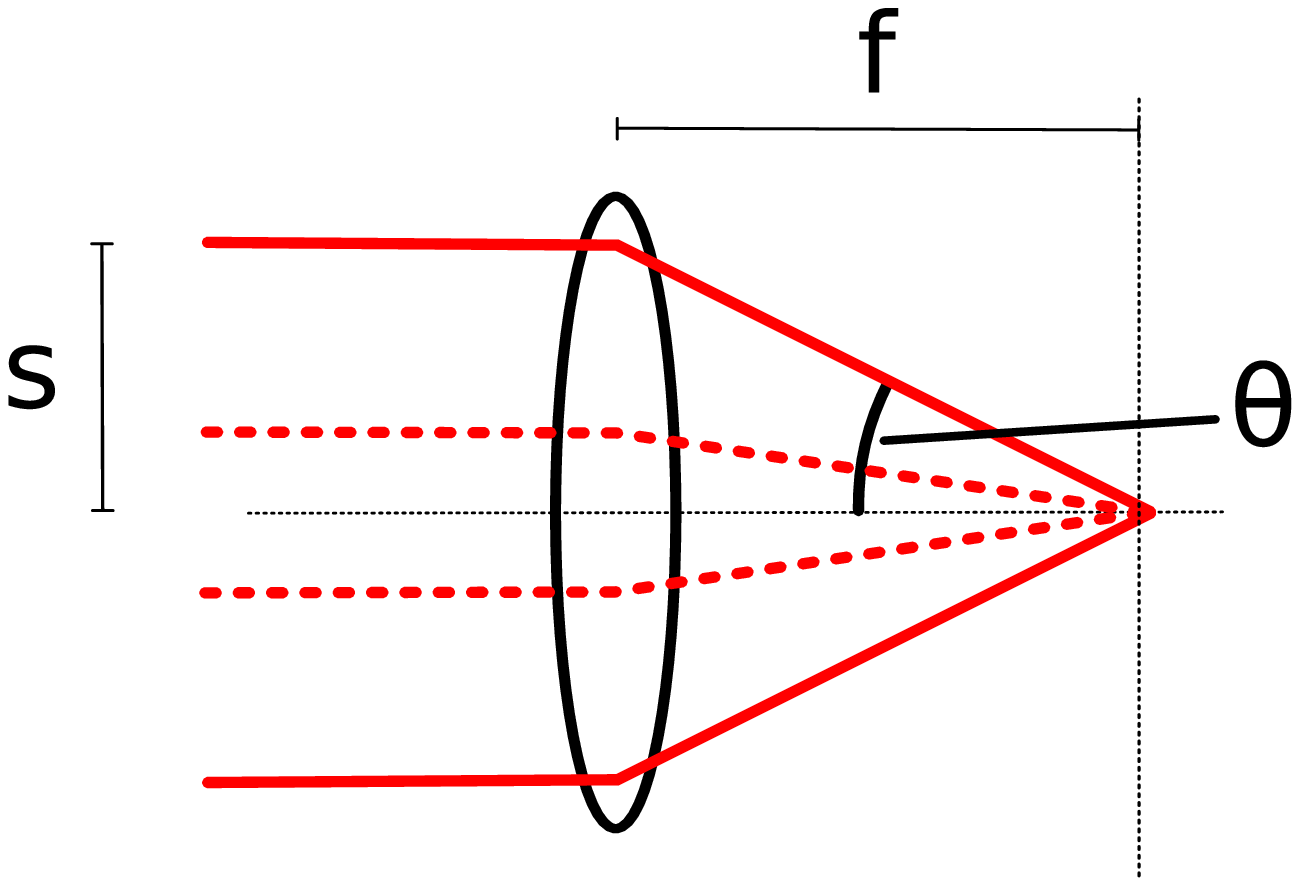}}\hspace{0.4cm}
\subfloat[]{\includegraphics[scale=0.3]{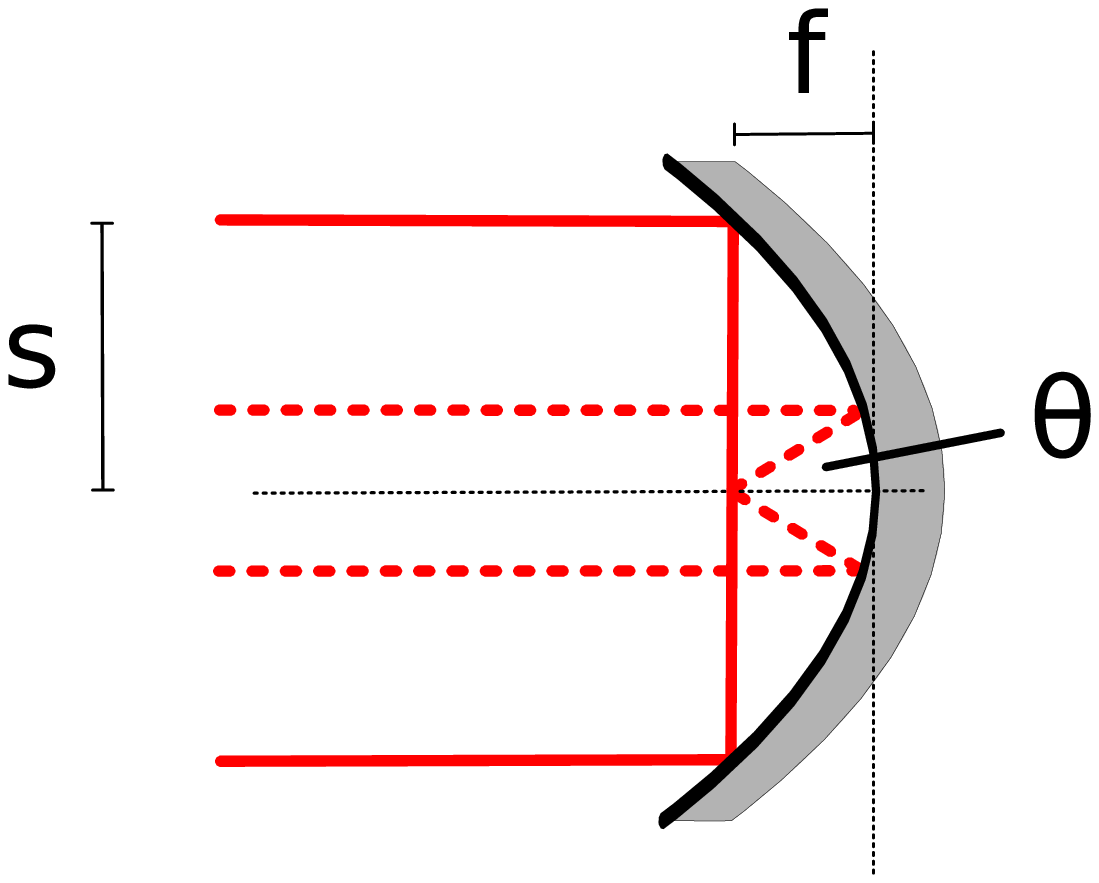}}
\caption{The basic geometries for the accordion lattice; only the last optical element is shown. The laser beams enter the last element parallel to the optical axis with an equal and opposite off-axis displacement which can be electro-optically controlled (see \cite{williams_dynamic_2008} for details). The lattice spatial period in either configuration is $d\,$=$\,\lambda/(2\sin\theta)$. Here $\theta$ is the angle between each laser beam and the optical axis at the position of the atoms; the maximum angle $\theta_{\textrm{\scriptsize max}}$ possible for the lens of Fig.\ (a) is given by the numerical aperture of the lens as sin$\,\theta_{\textrm{\scriptsize max}}$=$\,$NA.}\label{Opt1}
\end{figure}

\begin{figure}[t]
\includegraphics[scale=0.32]{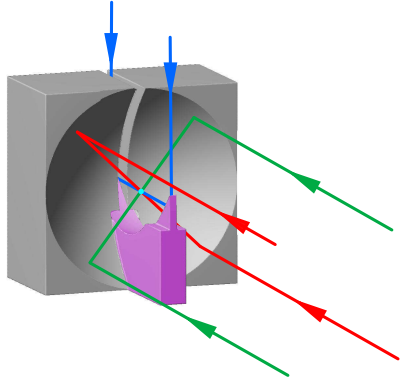}
\caption{(Color online) A mirror design for a three-dimensional parabolic counter-propagating accordion lattice. The underlying concept is to use a cylindrically-symmetric parabolic reflector (which has a central slice removed) to form a two-dimensional accordion lattice; a separate slim parabolic reflector is used to form the accordion lattice along the remaining dimension. Each pair of beams is generated separately. Unused parts of the parabolic reflector can be removed to improve the optical access. It is anticipated that the atoms would be observed using an objective lens (not shown) located in a diagonal position. The displayed design is not unique; variants are possible.}\label{Opt2}
\end{figure}

The accordion lattice is an optical lattice with a dynamically and smoothly variable spatial period \cite{li_real-time_2008,williams_dynamic_2008,alassam_ultracold_2010}. In Ref.\ \cite{williams_dynamic_2008} it is outlined how the lattice period can be controlled electro-optically using an acousto-optical deflector. The final optical element in the accordion lattice apparatus takes two beams, both parallel to the optical axis, directing them onto the atoms; the accordion lattice periodicity depends on the off-axis displacement $s$ of the beams. When the final optical element is a lens with numerical aperture NA (Fig.\ \ref{Opt1}(a)), the minimum lattice period is limited to $d_{\textrm{\scriptsize min}}\,$$=\,$$\lambda/(2 \, \textnormal{NA})$. Better performance is achieved using a parabolic mirror in place of the lens (Fig.\ \ref{Opt1}(b)); this reduces the minimum lattice period to the optimal $d_{\textrm{\scriptsize min}}\,$$=\,$$\lambda/2$. A three-dimensional `parabolic' accordion lattice can be formed using three independently controlled frequency-offset one-dimensional accordion lattices together with an appropriately constructed mirror element. One conceptual three-dimensional parabolic lattice mirror design is given in Fig.\ \ref{Opt2}.

The maximum spatial period of the accordion lattice formed using a parabolic mirror of focal length $f$ is $d_{\textrm{\scriptsize max}}\,$$=\,$$f \lambda / (2 s_{\textrm{\scriptsize min}}$). This is determined by the closest distance $s_{\textrm{\scriptsize min}}$ the incident beams can come to the optical axis (Fig.\ \ref{Opt1}(b)) without encountering a mechanical obstruction (e.g.\ see Fig.\ \ref{Opt2}) or overlapping with the atoms prior to reflection from the mirror.

In an experiment, the coherent quantum dynamics would occur while the lattice is in the minimum spatial period configuration, as both tunneling and interaction dynamics are strongest at small lattice periods. After the coherent dynamics have occurred, the lattice depth is increased to shut off interwell dynamics, pinning the atoms in place. The spatial period of the accordion lattice would then be smoothly increased for resolved-site measurements of the atomic occupation numbers.

The parabolic mirror (Fig.\ \ref{Opt1}(b)) both deflects and focuses the incident laser beams; collimated incident beams will be brought to a focus at the atoms. To compensate for this, the incident beams should have a focus before the mirror (on the front focal surface), so they are collimated at the atoms. However, the front focal surface of the parabolic mirror is curved, so this compensation can occur for only one off-axis displacement $s$. It is suggested that this compensation occurs for the minimum accordion lattice period $d_{\textrm{\scriptsize min}}\,$$=\,$$\lambda/2$, when the atoms undergo coherent dynamics. The maximum period lattice will still be somewhat distorted; however, this is unimportant, as coherent dynamics do not occur in the maximum period lattice.

\begin{figure}[t]
\includegraphics[scale=0.7]{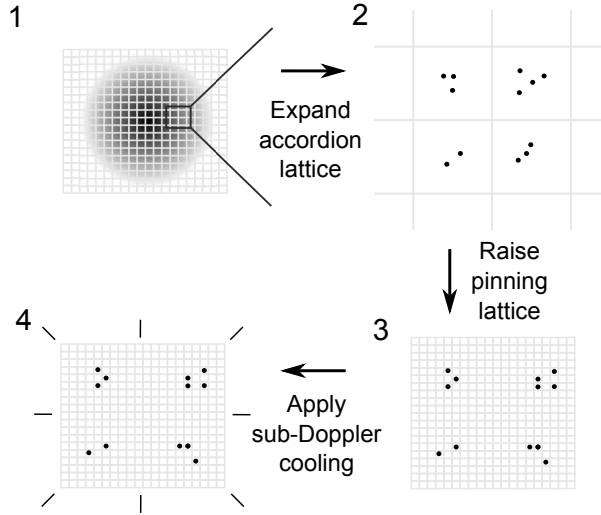}
\caption{Schematic overview of the accordion and pinning lattice measurement technique. The measurement sequence starts in the short-period accordion lattice (1) once the quantum dynamics of interest have taken place within the ultracold atomic sample (black). The accordion lattice height is increased, followed by the lattice period (2). The deep pinning lattice is raised (3); atoms in the same well in the accordion lattice are predominantly trapped in different wells in the pinning lattice. When the sub-Doppler cooling beams are turned on (4), the atoms fluoresce without being lost to light-assisted collisions. For certain pinning lattice and cooling beam parameters the mean lifetime of atoms at pinning sites can be many seconds \cite{shotter_large_2010}. While the individual pinning lattice sites cannot be resolved, atoms from different accordion lattice sites should be distinguishable using a microscope objective lens.}\label{transferProc3}
\end{figure}

Once the accordion lattice has been expanded to its maximum extent the lattice sites should be far enough apart to be resolved optically. Rather than directly illuminating the atoms in the accordion lattice, it is suggested that the atoms are at this stage loaded into an additional optical lattice---the pinning lattice. This is a deep three-dimensional counter-propagating optical lattice (with spatial period $\lambda/2$); as the intensity of this new lattice is increased, it divides each original accordion lattice well into many new wells; atoms are distributed between the new wells (Fig.~\ref{transferProc3}). The rate of the intensity increase of the pinning lattice should be fast enough that there is little transfer of atoms between the pinning lattice sites during the ramp-on of the pinning lattice; the distribution of atoms in the pinning lattice sites therefore has roughly the same envelope as the atomic density distribution in the accordion lattice before the pinning lattice was raised. After this (non-adiabatic) transfer process, the atoms will be trapped at a finite temperature in the pinning lattice. The accordion and pinning lattices do not need to be phase-locked. Once in the pinning lattice, atoms from the same accordion lattice site will be, to a high probability, isolated from each other. When the atoms are then probed with resonant light they cannot undergo light-assisted collisions; this addresses the third issue discussed in the introduction.  Note that the probing lattice can be much closer to the atomic resonance (e.g.\ a few thousand linewidths \cite{shotter_large_2010}) than the accordion lattice, which should be far-detuned to minimize light scattering during the coherent dynamics. 

The probability of mutually isolating the atoms depends on the ratio of the lattice site densities $(2d_{\textrm{\scriptsize max}}/\lambda)^{3}$ in the expanded accordion and pinning lattices; for example this ratio is $10^{3}$ for $d_{\textrm{\scriptsize max}}=5\lambda$. Defining $n_{a}$ as the number of atoms at an accordion lattice site, and $\langle n_{i} \rangle$ and $\langle \tilde{n}_{i} \rangle$ as the average number of atoms at pinning lattice site $i$ (which overlaps with the original accordion lattice site) before and after light-assisted collisions, it is found that 
\begin{equation}
\langle \tilde{n}_{i} \rangle=\sum_{{n}_{i}\text{ odd}}P(n_{i})=\frac{1}{2}-\frac{1}{2}\left(1-2\frac{\langle  n_{i} \rangle}{n_{a}}\right)^{n_{a}}\; .
\end{equation}
The fidelity of the division process is defined as the ratio of the average number of atoms observed at an accordion lattice site (after light-assisted collisions) to the true (original) number of atoms at that site; it is found to be
\begin{equation}
F=1-\frac{n_{a}-1}{n_{a}^{2}}\sum_{i}\langle n_{i} \rangle^{2}\;
\end{equation}
for $\langle n_{i} \rangle\,$$\ll$$\,1$. For example, assuming the atoms are at a temperature of $V_{0}/(4k_{B})$ in an accordion lattice with spatial period $d_{\textrm{\scriptsize max}}=5\lambda$ and potential barrier $V_{0}$ between sites (and that there is negligible transfer of atoms between pinning lattice sites during the ramp-on of the pinning lattice), the division fidelity is approximately 98.4\% for $n_{a}\,$=$\,2$, 96.9\% for $n_{a}\,$=$\,3$, and 95.3\% for $n_{a}\,$=$\,4$. Increasing the temperature of the atoms in the accordion lattice before loading into the pinning lattice actually improves the measurement fidelity, as the atoms are distributed between more pinning lattice sites.

Once the atoms are in separate sites of the pinning lattice they are subjected to polarization gradient cooling in three dimensions; the scattered light is collected to detect the atoms. The dynamics of this system, together with appropriate optical configurations and parameter ranges, are discussed in Ref.\ \cite{shotter_large_2010}. There, it is shown that a parameter range exists in which individual atoms are confined to separate pinning lattice sites, while scattering photons at rates of around $10^6\,$s$^{-1}$, until background gas collisions remove the atoms from the sample. To measure the three-dimensional density distribution of a sample tomographic measurements are required, with multiple images taken at different depths within the sample \cite{nelson_imaging_2007}. This can be achieved by translating the pinning lattice between images using electro-optical or piezoelectric means.

For one-dimensional, two-dimensional and thin three-dimensional samples, small clusters of atoms corresponding to the sites of the original accordion lattice will be resolvable using the above method. Notably, the method offers advantages over the method of the experiments described in Refs.\ \cite{bakr_probing_2010} and \cite{sherson_single_2010} by enabling the use of a smaller period ($\lambda/2$) lattice, which has a smaller site volume and so an enhanced non-linear interaction parameter during the coherent quantum dynamics; and by measurement of the true number of atoms at a site, rather than the parity of this number.

For thicker three-dimensional samples, the fourth issue discussed in the introduction to this article becomes important; if the fluorescent scatter from out-of-focus atoms is too great, the signal from atoms in the focal plane of the imaging system may be degraded or rendered indistinguishable from the background light in the image. Whether this is the case in a particular experiment depends on the density and thickness of the sample, the spatial period of the expanded accordion lattice, and the numerical aperture of the imaging system.

To suppress background light from out-of-focus atoms in thick three-dimensional samples an additional `probe' beam can be used. It is intended that the probing beam illuminates the sample in a thin sheet co-incident with the focal plane of the imaging system. The scattered probe light is collected to form the images, rather than the cooling light. Conveniently, the probe beam is resonant with a different atomic transition than the cooling and pinning lattice light, so that all light except that scattered from the probe beam is efficiently filtered out of the signal. Scattered light from out-of-focus atoms is therefore highly suppressed in the resulting images.

The probe beam changes the polarization gradient cooling dynamics of atoms in the pinning lattice. The effect of the probe beam is investigated using the hybrid Monte Carlo--Master Equation technique described in Ref.\ \cite{shotter_large_2010}. The simulation technique described in the reference is altered for the current situation by the addition of further laser beams and atomic states (with full orientational structure), to incorporate probe beam scattering and optical repumping dynamics. 

\begin{figure}[t]
\centering
\includegraphics[scale=0.58]{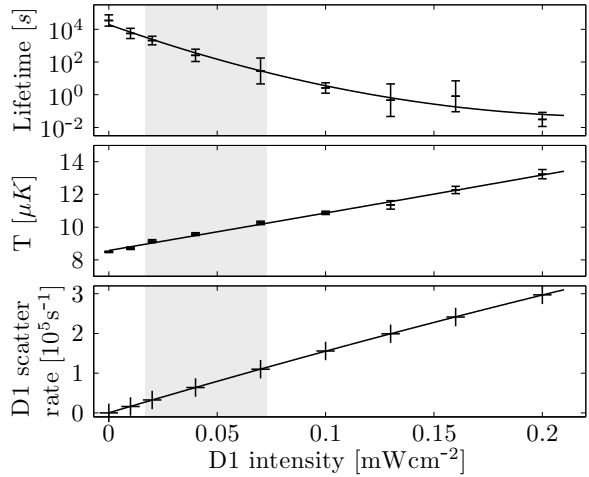}
\caption{Simulation of isolated $^{87}$Rb atoms in the pinning lattice undergoing polarization gradient cooling on the D$_{2}$ transition while being illuminated with probe light on the D$_{1}$ transition. The lifetime is the average time for an atom at a pinning lattice site (initially at equilibrium temperature) to leave that site (see Ref.\ \cite{shotter_large_2010} for details). The parameters used for the simulation are the same for the 1D alternating case in Table 1 of Ref.\ \cite{shotter_large_2010}, with the exception of the frequency and intensity of the pinning lattice light, which has a frequency detuning of 2000 linewidths to the blue of the F$\,=\,$2$\rightarrow\,$$3'$ D$_{2}$ line, and an intensity of 1.4$\,\times\,$10$^{4}\,$mWcm$^{-2}$ per beam; this gives the (bare) pinning lattice a depth of 410$\,$\textmu K. In addition, the simulation includes 30$\,$mWcm$^{-2}$ of repumping light resonant with the F$\,=\,$1$\rightarrow\,$$2'$ D$_{2}$ transition, and probing light in a linearly polarized beam resonant with the F$\,=\,$2$\rightarrow\,$$1'$ D$_{1}$ (795$\,$nm) transition. The error bars indicate the standard deviations of the quantities; these are due to the stochastic nature of the Monte Carlo simulation. The shaded region illustrates the approximate range of D$_{1}$ intensities which would be useful for the measurement technique described in this article.}\label{Perf}
\end{figure}

The simulation results are shown in Fig.\ \ref{Perf} as a function of the probe beam intensity. The probing light disrupts the polarization gradient cooling process by mixing and changing the relative energies of the ground state orientational sublevels; this is similar to the inhibition of polarization gradient cooling in a state-dependent lattice \cite{shotter_large_2010,winoto_laser_1999} or zero velocity cooling in a magnetic field \cite{walhout_sigma_1992}. Consequently the temperature of atoms in the pinning lattice increases, and the lifetime of an atom at a pinning lattice site decreases, with increased probe light intensity. If atoms leave their pinning lattice sites during the measurement, the measurement is degraded; this sets an upper limit on the useful probe beam intensity. It is seen from Fig.\ \ref{Perf} that a probe beam scattering rate of around 10$^{5}\,$s$^{-1}$ is possible with a mean site lifetime of order 100$\,$s; this lifetime is more than adequate for a probe beam exposure time of around 0.1$\,$s to 1$\,$s.

In order to demonstrate the potential of this technique it is worth illustrating these arguments with a concrete example of a possible lattice and probe beam geometry. It is assumed that a three-dimensional parabolic accordion lattice design is used with a mirror element similar to that shown in Fig.~\ref{Opt2}. The focal length of the parabola is assumed to be 8$\,$mm. The minimum distance between the accordion lattice beams $2 s_{\textrm{\scriptsize min}}$ is assumed to be 1.2$\,$mm (for the four horizontal incoming beams of Fig.~\ref{Opt2}) and 0.6$\,$mm (for the two vertical beams of Fig.~\ref{Opt2}), so that for a laser wavelength of 830$\,$nm the primitive cell of the expanded accordion lattice measures (5.5$\times$5.5$\times$11.1)$\,$\textmu m. It is advantageous to observe the lattice along a direction perpendicular to a diagonal lattice plane, as this increases the distance between lattice sites along the line of sight. Here it is assumed that the focal plane of the imaging system lies parallel to the ($\bar{1}12$) lattice plane (consequently the line of sight is accessible and lies in the horizontal plane). The projected distance in the focal plane between neighbouring accordion lattice sites is 4.5$\,$\textmu m and 3.9$\,$\textmu m in the horizontal and vertical directions respectively. The accordion lattice sites are therefore easily resolved with an imaging system which has a resolution of 1.3$\,$\textmu m (NA$\,$=$\,$0.35 at 795$\,$nm). It is clear that the pinning lattice sites are not resolved; however this is not required. The distance between the expanded accordion lattice sites along the line of sight is 19.2$\,$\textmu m. The probe beam sheet has a waist that is several hundred micrometres in the thick direction and 3.5$\,$\textmu m in the thin direction; this grows to 4.9$\,$\textmu m one Rayleigh length (48$\,$\textmu m) along the beam from the beam waist. The probe beam, which is parallel to the ($\bar{1}12$) lattice plane, therefore highly discriminates between different lattice sites along the line of sight. The probe beam alignment should be sufficient that the beam remains approximately co-incident with the focal plane of the imaging system across the width of the image; in comparison, the focal depth of a NA$\,$=$\,$0.35 system is around 5$\,$\textmu m at 795$\,$nm. Many exposures are taken as the pinning lattice is translated along the line of sight of the imaging system, with the exposures spaced by around half the probe beam waist (around 2$\,$\textmu m in this case); in this way the site-by-site three-dimensional distribution of atoms in the original accordion lattice can be reconstructed. 

In order to accurately reconstruct the atomic spatial distribution the fluorescence signal from each atom should have known amplitude. While predictable variations in the signal per atom, for example due to intensity envelope of the probe beam, may be known and compensated, unpredictable variations in the scattering rate should be minimized. 

One source of variability in the probe beam photon scattering rate is caused by differences in the relative phases of the cooling beams and the probe beam at the pinning lattice site. As discussed in Ref.\ \cite{shotter_large_2010}, changing the relative phases of the cooling beams at a lattice site can change both the cooling properties and the photon scattering rate of an atom at that lattice site. This is true even when a separate probe beam is used; changing the relative phases of the cooling beams alters the light field at a lattice site, which changes the atomic state occupancies and the probe beam scattering rate. For this reason it is suggested that the 1D alternating cooling beam configuration is used (see description in Ref.\ \cite{shotter_large_2010}) with an additional small frequency offset between the counter-propagating beams. The frequency offset rotates the net polarization vector of the (one-dimensional) cooling light at the lattice site relative to the probe light polarization; the scattering rate from the probe beam, when averaged over complete cycles, is now independent of all the phases of the beams at the lattice site. For the simulation of Fig.\ \ref{Perf}, a frequency difference of 3.6$\,$kHz was used between the counter-propagating beams (1/5 of the flash cycle rate), reducing the standard deviation in the scattering rate between lattice sites from 9\% (without this frequency offset) to negligible levels (less than 1\%).

Another source of variability in the scattered signal is due to the partially coherent nature of the scattered light. The coherent fraction of the total scattered light is \cite{cohen-tannoudji_atom-photon_2004}
\begin{equation}
\sum_{\epsilon}|\langle \langle \bm{D}^{+}_{\epsilon} \rangle \rangle|^{2} \big\slash
\sum_{\epsilon}\langle \langle \bm{D}^{+}_{\epsilon} \bm{D}^{-}_{\epsilon} \rangle \rangle \; ;
\end{equation}
here $\bm{D}_{\epsilon}^{\pm}$ is the atomic raising (lowering) operator, the index $\epsilon$ runs over \{$\sigma^{+}$,$\,\sigma^{-}$,$\,\pi$\}, and the doubled angled brackets denote time averaging of the expectation values. The coherent fraction is found to be 14\% for the simulations of Fig.\ \ref{Perf}. This is lower than for a two level atom under monochromatic illumination due to the multiple spontaneous emission routes and due to mixing of the atomic population between ground orientational substates by the cooling light. Interference between the coherent components of the scattered light from two or more unresolved atoms (i.e.\ atoms at the same original accordion lattice site) causes anisotropy in the scattered radiation and so unpredictable variations in the observed signal. These variations can be avoided by observing the atoms along the polarization vector of the probe light; this greatly diminishes the coherent proportion of the scattered light (for example, from 14\% to 1.3\% assuming an objective lens of NA$\,$=$\,$0.35). 

An unavoidable source of variation in the observed signal is the photon shot noise. To take an example, 100 photons are observed per atom during an exposure time of 0.1$\,$s for a scattering rate of 10$^{5}\,$s$^{-1}$, assuming 1\% of the scattered photons contribute to the observed signal; the shot noise is therefore 10\% for 1 atom, 6\% for 2 atoms, etc. This level of shot noise is small enough that 0 to 5 atoms per lattice site can be reliably counted (assuming other noise sources can be suppressed). A tomographic measurement of the atomic density of an extended sample may require between 20 to 40 images; the complete measurement would take 2$\,$s to 4$\,$s to complete at this frame rate.

The reabsorbed fraction of the scattered radiation depends on the optical depth of the sample $1/n\sigma$, where $n$ is the atomic density and $\sigma$ is the radiative cross-section. For example, a sample in an accordion lattice (with maximum spatial period (5$\times$5$\times$10)$\,$\textmu m) with an average of one atom per site, probed with the parameters used in Fig.\ \ref{Perf}, would have an optical depth of 8$\,$mm for the probe light and 15$\,$cm for the cooling light. These optical depths are large enough that the effects of reabsorption are negligible.

Although the technique is designed to measure the site occupancies of atoms in an optical lattice, it is also suitable for high resolution atom-by-atom measurements of any ultracold atomic sample which can be loaded into the accordion or pinning lattice (from, for example, a magnetic trap or free space) prior to the measurement without excessive distortion of the atomic density distribution. Site-selective preparation of a two-dimensional atomic ensemble would also be possible in the expanded accordion lattice, for example with use of an extra tightly-focussed photoionizing laser beam to eject atoms at particular accordion lattice sites. However it is probable that after this preparation process, extra in-lattice cooling would be needed for atoms in the short-period accordion lattice, in order to transfer them to the ground vibrational state at each site before coherent quantum tunneling dynamics take place.

In conclusion, it has been discussed how an accordion lattice can have a minimum spatial period of $\lambda/2$. When expanded, each accordion lattice site is optically resolvable. A separate pinning lattice can be used to mutually isolate atoms in the expanded accordion lattice, so that atoms cannot be lost due to light-assisted collisions during fluorescence imaging. It has been shown how a separate probing beam can be used to aid tomographic measurements of extended three-dimensional samples. Taken together, a technique is outlined which is capable of measuring the three-dimensional atom-by-atom site-resolved spatial distribution of ultracold atoms in a short spatial period ($\lambda/2$) optical lattice. 

The author would like to thank C.\ Foot (Oxford), the members of the Quantum Processes and Metrology Group (NIST Gaithersburg, USA), and the NIST internal referees; and to acknowledge funding from the EPSRC, Christ Church (Oxford), The Lindemann Trust and the Atomic Physics Division of NIST (Gaithersburg USA).

\end{document}